**Title**

# Basis for a hands-free blood flow measurement with automated vessel focus

**Authors**

R. Fuchs[1]*, N. Sumrah[2], J. Schwerdt[3,4], M. Unger[1], G. Stachel[2], M. Schultz[3], K. Lenk[2]†, T. Neumuth[1]†

[1]Innovation Center Computer Assisted Surgery (ICCAS), Faculty of Medicine, University of Leipzig, 04103, Leipzig, Germany
[2]Klinik und Poliklinik für Kardiologie, Universitätsklinikum Leipzig, 04103, Leipzig, Germany
[3]GAMPT mbH, Hallesche Straße 99F, 06217, Merseburg, Germany
[4]University of Applied Sciences Merseburg, 06217, Merseburg, Germany

*corresponding author - email: reinhard.fuchs@medizin.uni-leipzig.de
†contributed equally

**Abstract**

Cardiopulmonary resuscitation (CPR) is an essential tool to ensure oxygen supply during cardiac arrest, yet not quantifiable to this day. Low-quality chest compressions or wrong pressure placement go unnoticed. This paper presents a solution for the quantification of blood flow to guide first responders in their efforts. An approach for automated vessel identification with three different steps was developed, featuring a new sensor probe for ultrasonic measurements with non-symmetrically angled piezo ceramics. The probe was used with prototype ultrasound hardware for Pulsed Wave Doppler (PW Doppler) in a phantom. Initial measurements evaluated sensor vessel alignment at different sensor positions by examining Doppler results with a large sample volume during periodic flow. Afterward, an iterative mode was used for depth-dependent frequency measurements with score calculation of flow periodicity and power. The configuration with the best score was used for a prolonged monitoring mode. Initial mode and iterative mode aligned with ultrasound imaging regarding the best position and vessel depth. Simultaneous flow-sensor data and flow values of monitoring mode calculated via Doppler substitution showed a minimum correlation coefficient of 0.98, a minimum R2 of 0.96, and an average root mean square error of 3.84 ml/s. With the proposed hardware and software solutions, a basis for future developments was made, which could lead to a fully automated vessel identification and blood flow calculation during CPR. When used in emergencies, a miniaturized device could provide vital information about CPR efficiency that has yet to be included in the therapy of people during cardiac arrest.

# 1. Introduction

According to the European Registry of Cardiac Arrest (EuReCa) collaboration the annual incidence of out-of-hospital-cardiac arrest (OHCA) in Europe can reach up to 170 per 100,000 inhabitants. Resuscitation is attempted or continued by emergency medical services (EMS) personnel in about 50 % to 60 % of cases, while the rate of bystander CPR is 58 % on average. Survival rates at hospital discharge amount to an average of 8 %, varying from 0% to 18% [1]. Inadequate CPR is one of the main reasons for an unsatisfactory outcome after return of spontaneous circulation (ROSC). Guidelines for frequency and depth of chest compressions exist, but are difficult to verify and do not guarantee adequate oxygen supply to the brain via blood flow [2]. Cerebral blood flow itself appears to be a decisive factor for long term survival following a delayed ROSC, but measuring hemodynamic changes with invasive procedures is infeasible during resuscitation [3, 4]. A direct, non-invasive, and easy-to-use measurement for the out-of-hospital setting is not yet available [5, 6]. This hindered any conclusive research regarding the relationship between measured blood flow during CPR and neurological outcome. In addition, decisions regarding the start or continuation of invasive therapies in the intensive care unit are often pending until several days after resuscitation, to be able to perform a reliable neurological prognosis [7].

Emergency and intensive care can therefore profit from a tool to automatically detect and quantify blood supply of the common carotid artery (CCA). The modality could inform first responders about the effect of their measures, while records about no flow or low flow phases might provide insights about prognostic estimations of these patients [3, 4, 8-17]. Non-invasive blood flow measurements during resuscitation are technically possible using commercially available ultrasound (US) devices, but so far this could not be transferred to the prehospital situation with the technology and expertise currently available [8, 9]. While portable ultrasound devices with integrated Doppler frequency analysis exist (e.g. Clarius L7 Linear Handheld Ultrasound Scanner, Siemens ACUSON Freestyle), they do not have mounting options for robust hands-free attachment to the neck during CPR [18, 19]. Users also require expertise in device parameterization and signal interpretation for an explicit analysis of flow values inside the target vessel, which is not guaranteed in emergency services. The current market situation therefore offers no possibilities for non-invasive measurement of the blood flow of the CCA without extensive manual labor that would impede therapy measures.

Multiple solutions for a non-invasive, hands-free method are object of research and development by different research groups, many of them focusing on Doppler ultrasound analysis via a sensor fixated by adhesive materials [10, 12, 13, 15-17, 20]. Kenny et al. presented their solution *FloPatch* (FloSonics, Sudbury, Ontario, Canada), comprising of an ultrasonic patch employing continuous wave (CW) Doppler and two transducer arrays to detect any fluid movement below the surface the probe is attached to [12, 20]. The patch can be applied to the skin via adhesive strips, enables a hands-free measurement during any resuscitation actions and features automated algorithms for a quantified flow-measurement at the chosen position. Faldaas et al. focused on a similar approach, with a similar patch-based device called *DopplerRescue* (study device from AHL center, St. Olav's hospital, Trondheim, Norway), which uses a two-transducer setup employing pulsed-wave (PW) Doppler and is applied with adhesives as well [13]. The PW Doppler enables the user to select certain sample volumes (SV), i.e. specific times during which the transducers of the probes start and stop recording ultrasound reflections. Available tissue depths of *RescueDoppler* reach from 8 mm to 45 mm in 32 different steps.

However, an exact identification of particle movement inside the vessel is not possible with CW Doppler. The continuous application of ultrasound with one transducer array and the continuous measurement with the other leads to an increase in quantification accuracies due to the large tissue coverage. If any vessel direction changes occur inside the SV, the resulting measurements are corrupted, since the angle cannot be accounted for and the SV cannot be targeted to definitive depths. The system provided by Faldaas et al. allows for this feature due to PW Doppler. Yet the system needs manual adjustment, impeding usability in emergency scenarios with little time. It requires in-depth knowledge of the patients' anatomy and the skill to perform system adjustments. Since available parameters are pre-set, the possibilities for adjustment might not cover the necessary measurement setup for an SV inside the target vessel. Without information about potential or missing overlap between the vessel and focus point, an automated calculation correction is impossible. Other developments in the field of ultrasonic technology, material science, and microelectronics enable multiple research groups to develop flexible US patches that target similar issues; however, they suffer from similar problems [14-17].

This paper presents methods and materials that aim to address the issue of automated vessel identification and flow quantification using PW Doppler for iterative sample volume optimization. The described methods are evaluated for the possibilities to identify valuable Doppler positions and measure blood in the CCA or similar vessels, without the need for any adjustments regarding measurement parameters. The developed solution should be usable for manual detection of SV-coverage of the common carotid artery. A correct positioning of the US probe should allow a depth determination via iterative shifting of small SV. Multiple piezo-ceramic arrays should be usable for automatic measurement parameterization while the ceramics' fixed angle relationships should be used for approximation of the Doppler angle and the resulting blood-flow values.

# 2. Material and Methods

## System setup

Through our research and further dialogs with medical experts of the University Hospital of Leipzig, we propose a system usable for automated measurement of blood-flow in the CCA during resuscitation scenarios that should provide the following features:

- The device needs a quick method to check for vessel coverage inside a large SV.
- If found, the device needs to adjust the SV until it covers 2-4mm inside the vessel.
- The Device needs multiple transducers with PW Doppler for enabling multiple Doppler angles.
- The device should automatically calculate the flow velocity.
- The device should require a short customization time, preferable close to 5 seconds [2].

The final demonstrator system consists of a computer and a sensor probe with three differently angled piezo ceramics. The ultrasonic angles result of the ceramics' orientation and are 25°, 0°, and -12° toward the probe surfaces' normal vector. The casing of the probe is made of polyactide and the filling along the path of the soundwaves is polyetheretherketon (PEEK) for ultrasound propagation. The probe is connected to a main computing unit via three ultrasound modules, one for each ceramic. Synchronized communication enables the main unit to trigger and measure ultrasonic pulses independently in all ceramics. Fig. 1 shows a schematic concept of the probe as a crosscut above tissue and a fluid-filled vessel with active flow.

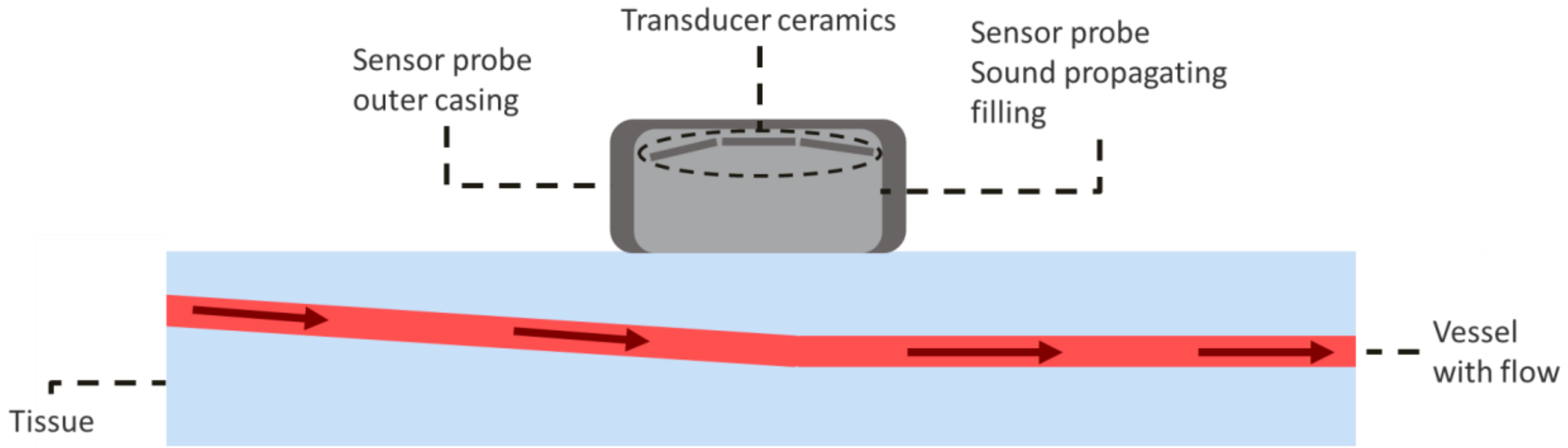

Fig. 1 Concept drawing of the internal configuration of the US-probe of the system with underlying tissue and vessel cavity

## Laboratory setup

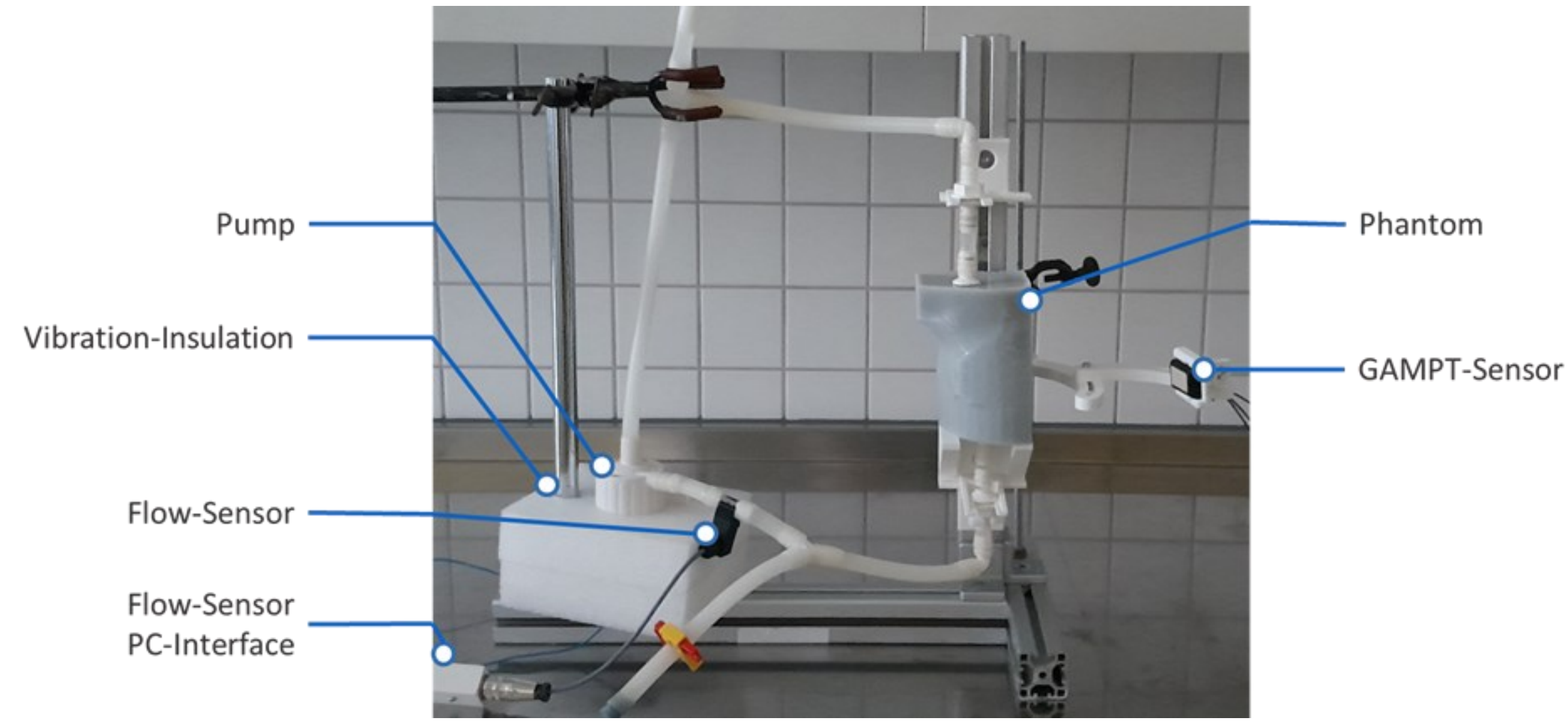

Fig. 2 Laboratory setup for simulated flow measurements featuring GAMPT mbH pump and phantom

For evaluation of the automated measurement process in the laboratory, a neck phantom was created by following the standard operating procedure to prepare an agar phantom [21]. The agar-agar mixture was used to fill a 3D-printed mold of a partial neck, which included a 3D-printed vessel with 8 mm diameter. Leftovers of the mixture were used to create small sample blocks for verification of soundwave propagation behavior. The phantom was connected to a mechanical pump and silicone tubes, and the resulting circulatory system was filled with Doppler fluid. The pump and the Doppler fluid were provided by GAMPT mbH (Merseburg, Germany). Blood flow itself was simulated by automated pulsed activation of the pump with 60 BPM. For reference, the periodical flow was measured with a VISION 1005 2F66 flow-sensor (Badger Meter, Wisconsin, United States). The overall laboratory setup for the measurements is shown in Fig. 2.

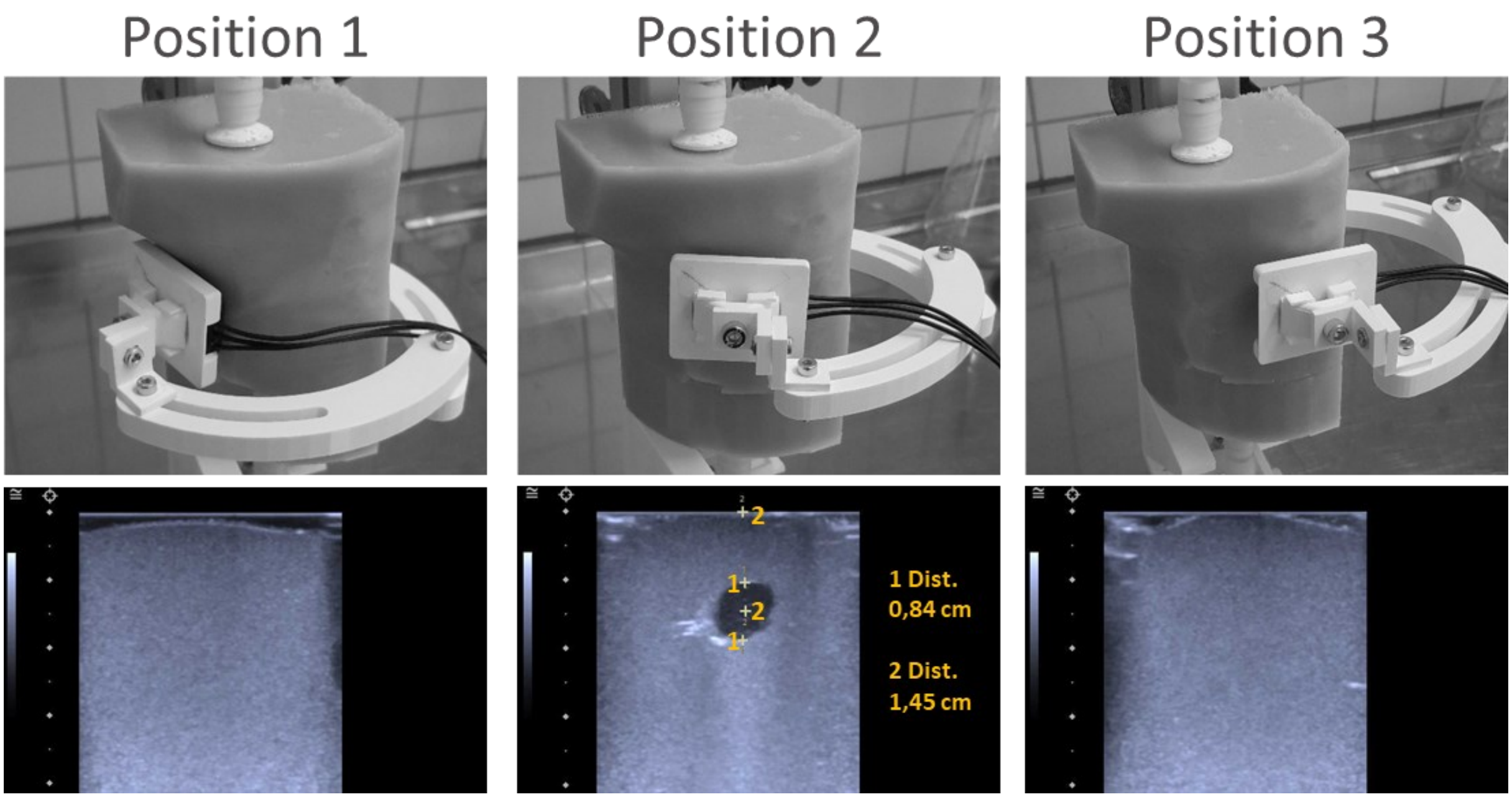

Fig. 3 Measurement positions and respective B-Mode images for vessel coverage validation; Left/Position 1 & Right/Position 3: proximal and medial placement with no coverage; Middle/Position 2: middle placement with coverage of the vessel, diameter (1 Dist.) and depth (2 Dist.) measurements.

For flow measurements, the Doppler system probe was placed at three different phantom locations, displayed in Fig. 3 along with B-Mode images taken before the actual measurements. For repeatability and reproducibility, the probe was placed by using a specially designed probe holder that would ensure the probes position during the measurement. Ultrasound gel was used for coupling of the transducer to the phantom. Once the sensor was positioned, different measurement modes (initial, iterative, and monitoring) of the new Doppler system were applied in succession. The initial mode was used at every position. If no blood flow could be verified via real-time visualization, no further measurements were executed. B-mode imaging was done with the ACUSON Sequioa (Siemens Healthineers AG, Forchheim, Germany).

### Measurements – Initial mode

The initial mode of the system uses the transducer with 25° to transmit ultrasound into the underlying tissue every 50 µs. Each transducer module was set to collect reflecting soundwaves after 5 µs with a measurement duration of 30 µs, i.e. the gate start and size of the sample volume (SV). The parameters were calculated with equation (1), where $s$ marks the traveled distance, $t$ the time and $c$ the ultrasound propagation velocity inside the tissue. Distances for gate start and size were chosen empirically. Their values equaled 3.5 mm and 20 mm respectively. Resulting times were rounded up. Tissue specific sound propagation of the phantom was approximately 1540 m/s, which was verified by comparing manual dimension measurements with B-Mode results done on the agar blocks.

$$s = \frac{t}{2} * c \tag{1}$$

Through this mode, an initial measurement is made over a big SV. If no blood flow occurs in the area, the sensor position is considered unsuitable for a more exact analysis and the sensor is moved to another position, where the initial mode is repeated. If blood flow is occurring in the sampled tissue volume, the characteristic Doppler spectrograms can be observed via either the monitor of the system or the generated sound of the received Doppler frequencies by the system. If the periodical flow was observable, the position was used for the iterative mode afterwards.

### Measurements – Iterative mode

The iterative mode employs smaller receiving windows for the Doppler measurements at multiple tissue depths. The system uses only one transducer as transmitter, yet all ceramics are employed for Doppler measurement. The mode started with an SV that contained the sound reflections measurable after 5 µs following the ultrasound burst. The measurement persists for 2 µs. After reconstruction of the sampled Doppler-frequencies at that particular tissue depth, the delay between burst and measurement was increased by 2 µs. Stepwise increase of the measured tissue depth was repeated until the SV reached a starting time of 43 µs. Once the complete range of SVs had been measured and the data was analyzed, the transmitting transducer was switched and the measurement loop started again. This was done until all three transducers served as ultrasound transmitter. Finally, analysis results were compared to find the transducer whose ultrasonic emission allowed for Doppler measurements with the most distinctive blood-flow curves.

### Measurements - Monitoring mode

Following the identification of the best transmitter and the depth with the highest score an additional measurement was performed at that SV with 2 µs range. Volume curve calculation resulted in three transducer curves and one flow-sensor curve, which were recorded during pulsed pump activation. Pulse rate was 60 BPM and peak flow during the pulses was adjusted along the time, first increasing peak flow values step-wise and then decreasing step-wise.

## Analysis – Initial mode

All collected data were analyzed with the software Matlab v2024a (The MathWorks, Inc., Natick, Massachusetts, United States). Before acquisition by the ADCs, the signals were split into I and Q components, which were used for spectrogram calculation after offset removal. The Doppler data were used for spectrogram reconstruction $P_s(t,f)$ and envelope calculation $g_{MGM}(t)$. Spectrogram calculation was done with the MATLAB spectrogram algorithm. With the function, the short-time Fourier transformation (STFT) for each ultrasound pulse reflection was calculated with a frame length of 1024 samples, an overlap of 512 samples, 4096 values, and a sample frequency of 20 kHz. A lower power threshold was applied to eliminate background noise and small artifacts. The threshold value was 10e+03 arbitrary units (a.u.), based on the maximal pixel values in the noise region of the Doppler spectrogram with a factor of ten applied. The noise region was defined as the frequency range above 5 kHz. The resulting power spectrums of each burst were concatenated to form the spectrogram $P_s(t,f)$ of one SV measurement, which was finally used for the calculation of the Doppler shift envelope curve $g_{MGM}(t)$. Envelope calculation was performed with the Modified Geometric Method (MGM) for every spectrogram [22]. For the initial mode and the monitoring mode, the resulting envelope curves were divided by two to account for inhomogeneous inner vessel flow velocity along the vessel diameter. Each curve was filtered with a moving average filter with a window size of 5 samples.

The synchronized flow-sensor data, which had been recorded with 10 samples per second (SPS), was converted from number of rotations per 0.1 s to ml/0.1 s in accordance with sensor documentation. The resulting volume per 0.1 s was divided by the circular area, calculated via the inner diameter of the sensor ($d_{sensor}$ = 6 mm, $A_{sensor}$ = 1.13 cm$^2$), to calculate the flow velocity $v$ inside the sensor.

Temporal similarity of the flow sensor and velocity curves was analyzed with the coefficient of determination $R^2$ and the correlation coefficient. For $R^2$ calculation, the MATLAB proprietary algorithm *fitlm* was used. The function was employed for a linear fitting method between the ground truth curve and each transducer curve. The correlation coefficient calculation was done with *corrcoef*, which calculates the Pearson correlation coefficient.

## Analysis - Iterative mode

For the evaluation of overlap between SV and the vessel center during the iterative mode, the envelope $g_{MGM}(t)$ of each spectrogram was analyzed in the frequency domain by using the Fourier analysis. These envelope curves were used without scalar division by two. Following the transformation via the discrete Fourier transformation to $G_{MGM}(f)$, the analysis algorithm calculated the frequency with the highest signal power in the desired frequency range. The range was between 42 BPM and the optimal cardiopulmonary resuscitation frequency, estimated at 120 BPM [2].

After detecting the signal frequency with the highest power in between the frequency thresholds, i.e., 42 BPM / 60 s = 0.7 Hz and 120 BPM / 60 s = 2 Hz, harmonic frequencies were calculated by multiplying the detected frequency $f_{max}$ by $i = [1,2,3]$. The power of the three frequencies was summed up for the power estimation of the pulse signal. To account for instability in the periodic flow and the distribution of the power around the harmonic frequencies, adjacent frequency powers were added. The signal power was located at frequencies $f = i * f_{max} + j * f_d$ with $j = [-1,0,1]$ and $f_d$ being the frequency resolution of $G_{MGM}(f)$. The final sum was divided by the sum of $G_{MGM}(f)$ located in frequencies between 0 Hz and 20 Hz, resulting in $P_G$, shown in (2).

The envelope of each five-second segment was further used to calculate the sum of all power values $P_s(t,f)$ of Doppler frequencies that lie below $g_{MGM}(t)$, i.e., the mark of the end of the power bulk of the STFT, shown in (3). Each sum $P_{MGM}$ of each segment was multiplied by the calculated $P_G$ value, combining the power of ultrasonic reflections and the periodicity of the extracted flow signal. The final parameter was defined as the MGM power score $P_{score}$. This power score was calculated for each five-second measurement, executed by each transducer at each respective SV, according to (4).

$$P_G = \frac{\sum_{i=1}^{3}\sum_{j=-1}^{1} G(i * f_{max} + j * f_d)}{\sum_{f=0\,Hz}^{20\,Hz} G(f)} \tag{2}$$

$$P_{MGM} = \sum_{t=0\,s}^{5\,s} \sum_{f=0\,Hz}^{g_{MGM}(t)} P_s(t,f) \tag{3}$$

$$P_{Score} = P_G * P_{MGM} \tag{4}$$

All scores were used for the determination of the transducer whose ultrasonic emission led to the maximal score within the multiple SV loops and the respective depth. For this experiment, the overall maximum power score was determined to acquire the best possible ultrasonic emission transducer at that sensor position. Afterward, the maximal power score of the three receiving transducers along the SVs was chosen.

### Analysis – Monitoring mode

The recorded flow sensor data was converted from pulse/0.1 s to ml/0.1 s and interpolated to the sample frequency of the envelope curve $g_{MGM}(t)$, 40 SPS. Afterward, the volume curve was filtered with a 40-point moving sum. Finally, scalar multiplication with a factor of 0.25 achieved the average volume per 1 s. To attain the velocity and volume curves of all three transducers, spectrogram and curve calculations were performed as described in the analysis of the initial measuring mode. Resulting curves were used for cross-calculation to substitute the unknown Doppler angle, i.e., the angular relation between the ultrasound direction from the sending transducer and the flow direction inside the vessel, with known fixed transducer angles, as it is described in Supplementary Equations S1. Doppler angle substitution was done three times, once each time between transducer (T) pairs T1 and T2, T2 and T3, and finally T3 and T1. Velocity $v$ was converted into a volume flow $\dot{V}$ via (5), employing the measured vessel radius $r$ of the B-mode image of the phantom (see Fig. 4), assuming an approximate circular shape.

$$\dot{V} = v * \pi * r^2 \tag{5}$$

Finally, moving sum filtration was applied to the volume curves in the same way as it was done to the flow sensor curve, to receive curves in ml/1 s. The transducer curves and flow-sensor data were analyzed for correlation via $R^2$, the correlation coefficient, and the root mean square error (RMSE). $R^2$-calculation was done with the Matlab proprietary algorithm *fitlm*, using a linear fitting method between the ground truth curve and each transducer pair curve. For correlation coefficient calculation, the function *corrcoef* was employed, which calculates the Pearson correlation coefficient. RMSE calculation was done with *rmse*.

## 3. Results

### Measurements – Initial mode

Initial measurements with the wide SV on three different positions resulted in two failed and one successful vessel observation by the new Doppler system. Vessel coverage by the SV was verified by visual comparison of the reconstructed spectrogram, the calculated envelope curve, and the simultaneously measured flow-sensor data. Fig. 4 shows the three measurement positions with the resulting spectrogram and envelope curves for each transducer. The transducer plots start at the top with transducer 1 and end at the bottom with transducer 3. Additionally, each graph includes the flow-sensor data, which serves as ground truth. Position 1 and position 3 show no significant ultrasound reflections in any of the transducer signals and have therefore an envelope that equals zero. To save space both positions are represented with the same graphs. Position 2 shows strong ultrasound reflections in each of its transducer data and envelopes seem to correlate with the flow-sensor curve in the time domain.

## Measurements – Iterative mode

After the initial measurements, the stepwise scan of the tissue via iterative measurements with small windows and step sizes was done for position 2. Starting with an SV between 5 µs and 7 µs (equals approx. 3.75 mm and 5.25 mm), the iterative mode repeated measurements for 20 successive steps, resulting in 21 measurements for each transducer per ultrasonic emission-measurement-regime. All in all, this resulted in 189 spectrograms for analysis.

Fig. 5 shows selected spectrogram results with their approximated SV origin by connecting them to an area on the B-Mode image for position 2. The corresponding measurements originate from ceramic 1 (25° angle) with ceramic 3 (-12° angle) as acting sender and selected steps are the SV between 15 µs and 33 µs. The reconstructed spectrograms are shown on the right of the position 2 ultrasound image overlaid with a grid. Each crescent area of the grid correlated to one SV and is connected to its respective graph. The axis on the left is divided into 0.5 cm steps, with bigger dots marking 1 cm steps. The axis starts with 0 cm and ends with 3 cm. The spectrograms are listed from left to right and from top to bottom in increasing SV depth. Therefore, the spectrogram in the upper left corner contains the data that is closest to the phantom surface and the lower right corner shows the data selection with the farthest SV. It can be seen that the power of ultrasonic scattering signal starts to occur at the SV between 19 µs and 21 µs and peaks during 21 µs to 23µs and 23 µs to 25 µs. According to formula (1) this equals to 15.75 mm and 18.75 mm tissue depth. Following their strongest flow representation in those two SV, the following spectrograms decrease in their overall power until there is no discernable power contained in the reconstruction at SV 31 µs to 33 µs. The inner distance of the probe between transducers and the sensor surface is approximately 3 mm, which needs to be subtracted from the measurement results, to allow for a comparison with the B-mode image.

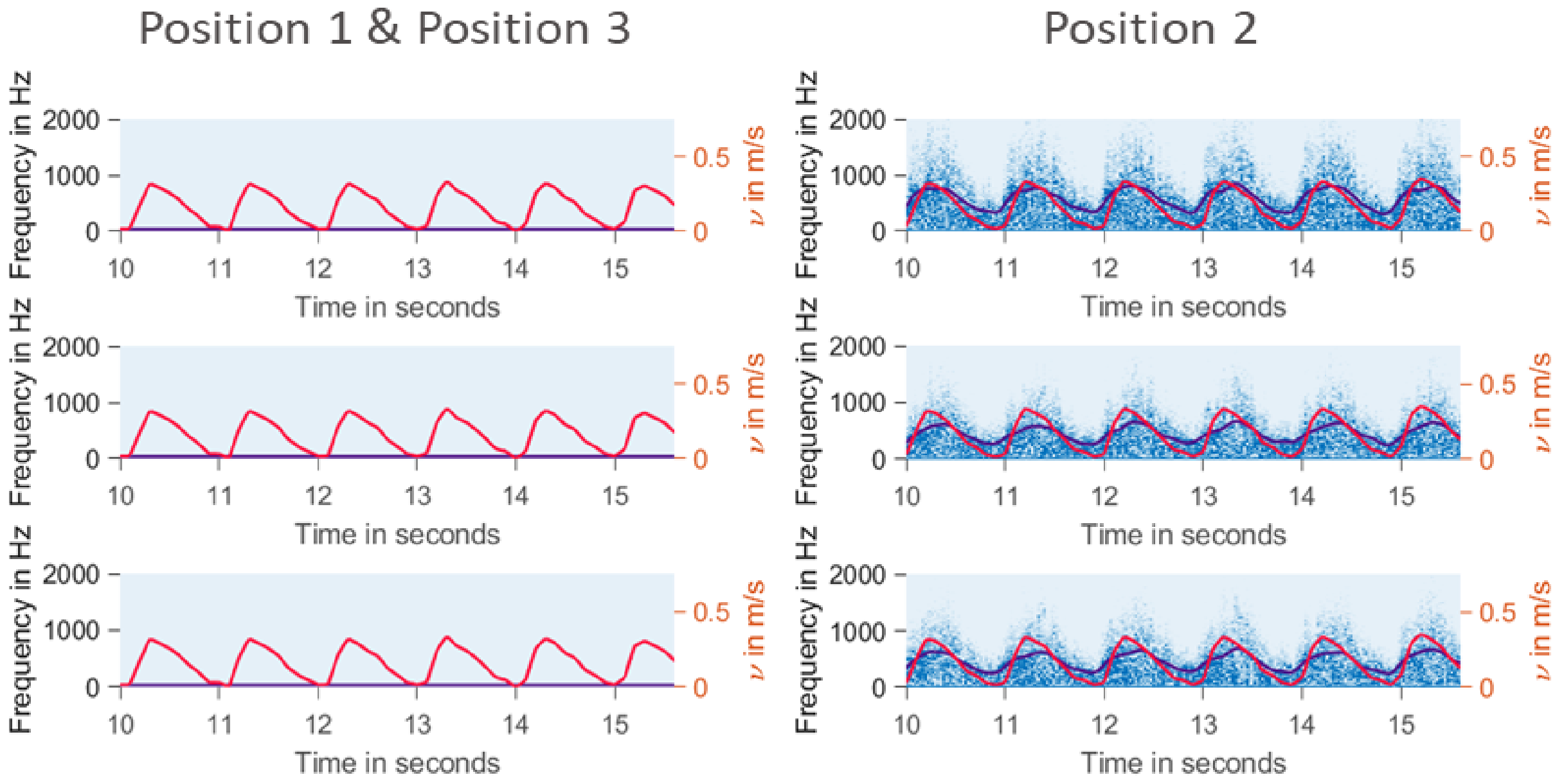

Fig. 4 Overview of transducer spectrograms, envelopes (blue) and Flow-Sensor data (red) at the three positions showing the correlation of measured Doppler data and the flow reference; Position 1 & 3: no apparent US reflections during active flow in all three transducer Doppler spectrograms; Position 2: Flow curve and coincidental Doppler data of recorded reflections

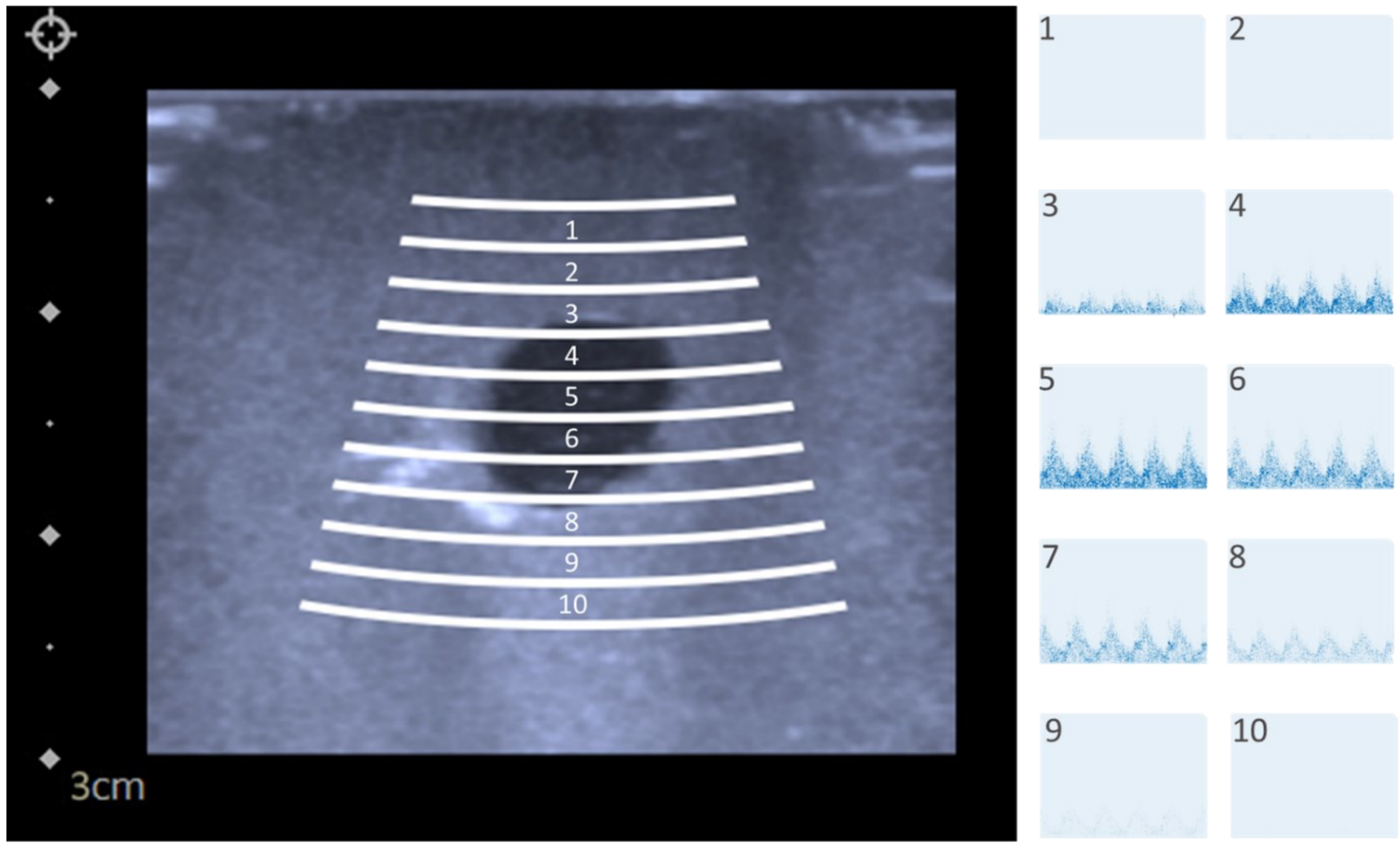

Fig. 5 Iterative mode results showing the increase and decrease of Doppler signal power when sample volume overlaps and misses vessel cavity; Left: B-Mode image showing the vessel cavity and a grid denoting the step-wise sample volume displacement concept; Right: Results of ten different measurement depths displayed as reconstructed spectrograms

## Measurements – Monitoring mode

Measurements with the small SV were done during different peak velocities with varying time frames. After no flow, the peak velocity was increased three times, before reducing it three times. During the first reduction, a manual error resulted in a short negative peak in the control signal, which can be seen in Fig. 6 at t = 100 s after the first flow decrease. The figure shows three volume curves, resulting from the combination of two transducer curves each and the reference, the flow-sensor signal, converted to ml/s. All curves cover a similar value range and show synchronized behavior during moments of flow increase and decrease. The end of the measurement shows no-flow. However, all transducer signals have an offset, which is due to the vessel filter that cuts of frequency range at 75 Hz during envelope calculation.

## Analysis – Initial mode

Results of correlation tests of each transducer curve with flow-sensor data as ground truth are shown in Table I. Each comparison shows that all results lie above 0.6, indicating a certain amount of similarity. While $R^2$ results reach an average of 0.70, the average correlation coefficient is 0.83.

**TABLE I CORRELATION OF SPECTROGRAM ENVELOPE OF INITIAL MEASUREMENT WITH FLOW-SENSOR AS REFERENCE**

| Target curve | $R^2$ | Correlation coefficient |
|---|---|---|
| T1 | 0.75 | 0.87 |
| T2 | 0.64 | 0.80 |
| T3 | 0.70 | 0.83 |
| Average | 0.70 | 0.83 |
| Standard deviation | 0.06 | 0.03 |

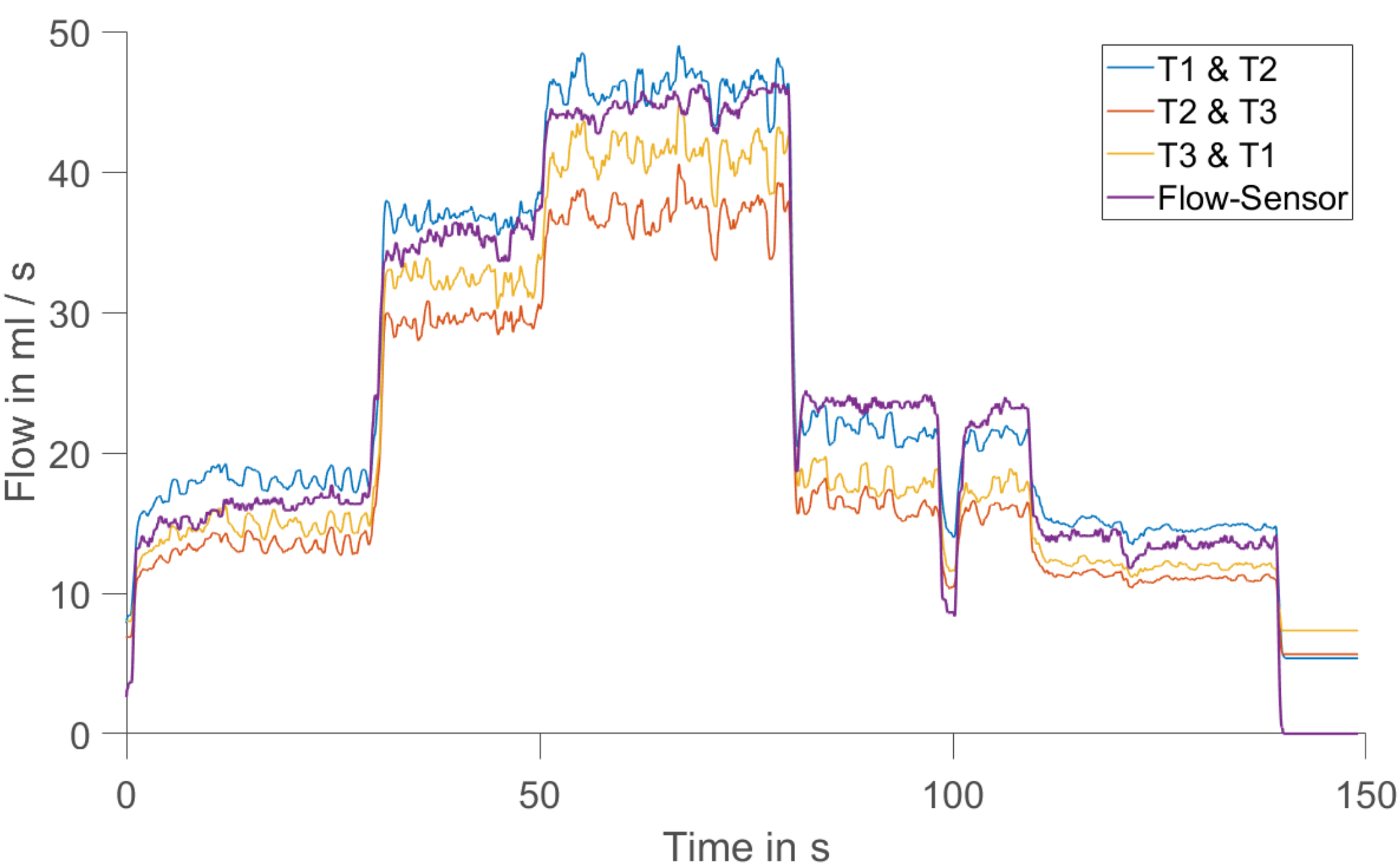

Fig. 6 Overlapping volume flow curve results from Doppler flow calculation and the flow-sensor signal showing qualitative and quantitative differences during different pump settings, both as volume per second curves with transducer volume curves derived from combined transducer velocity curves

## Analysis – Iterative mode

Each stepwise scan during iterative analysis resulted in a five-second window that allowed for a power evaluation in the frequency spectrum (after detrending and Hanning-Window application), via summation of the most powerful frequency between 0.7 Hz and 2 Hz as well as its two harmonics and the division by the total power and multiplying with the sum of frequency powers below the envelope. Results were divided by a factor of 1e+08 a.u. for scaling. The results of these calculations are gathered in Table II, showing the corresponding score for each SV, which is listed as starting time in µs and corresponding depth in mm. The table lists the receiving score under the corresponding receiver and the responsible US sender. Comparing the resulting power metric, describing the power and periodicity of Doppler reflections, the maximal value of 12.2e+08 a.u. was found at SV 21 µs (15.75 mm) when transducer 2 acts as the sender and transducer 1 as the receiver.

**Table II Score results of step-wise Doppler measurement during flow with 60 BPM pulse divided by 1e+08 a.u.**

| SV | | Sender Transducer 1 | | | Sender Transducer 2 | | | Sender Transducer 3 | | |
|---|---|---|---|---|---|---|---|---|---|---|
| µs | mm | R T1 | R T2 | R T3 | R T1 | R T2 | R T3 | R T1 | R T2 | R T3 |
| <17 | <12.8 | 0.3 | 0.0 | 0.0 | 0.0 | 0.0 | 0.0 | 0.0 | 0.0 | 0.0 |
| 17 | 12.8 | 0.3 | 0.0 | 0.0 | 0.0 | 0.0 | 0.0 | 0.0 | 0.0 | 0.0 |
| 19 | 14.3 | 3.9 | 2.9 | 1.1 | 8.2 | 2.8 | 1.2 | 2.6 | 1.1 | 1.7 |
| 21 | 15.8 | 6.3 | 6.1 | 5.1 | 12.2 | 5.7 | 3.6 | 8.8 | 3.3 | 4.6 |
| 23 | 17.3 | 7.8 | 6.1 | 6.4 | 11.3 | 5.7 | 4.1 | 11.7 | 4.0 | 4.8 |
| 25 | 18.8 | 6.8 | 3.9 | 5.0 | 7.3 | 5.0 | 4.1 | 9.4 | 4.1 | 4.2 |
| 27 | 20.3 | 4.9 | 2.3 | 2.5 | 4.4 | 3.4 | 2.6 | 6.4 | 2.9 | 2.5 |
| 29 | 21.8 | 0.0 | 0.0 | 0.0 | 2.2 | 2.3 | 1.7 | 3.0 | 1.8 | 1.4 |
| 31 | 23.3 | 2.9 | 0.0 | 0.0 | 0.7 | 0.8 | 0.2 | 0.7 | 0.5 | 0.7 |
| >31 | >23.3 | 0.1 | 0.0 | 0.0 | 0.0 | 0.0 | 0.0 | 0.0 | 0.0 | 0.0 |

### Analysis – Monitoring mode

Table III shows the results of the $R^2$, the correlation coefficient calculation, and the mean value as well as the standard variation of the scalar curve differences between the calculated flow-sensor volume curve and the combined transducer measurements at SV 21 µs to 23 µs. Rounded $R^2$ values for the three comparisons range from a minimum of 0.96 (combination of transducer 2 and transducer 3 as well as transducer 3 and transducer 1) to a maximum of 0.98 (combination of transducer 1 and transducer 2), with a rounded average and standard deviation of 0.97 and 0.01, respectively. The correlation coefficients reached a value of 0.99 for the combination of transducer 1 and transducer 2. The remaining combinations have a correlation coefficient of 0.98, which equals the average with a standard deviation of 0.006 before rounding. RMSE of the curves showed a difference between the flow-sensor and combinations T1T2, T2T3 and T3T1 of 2.31 ml/s, 5.50 ml/s and 3.72 ml/s respectively. Average RMSE was 3.84 ml/s and their standard deviation 1.59 ml/s.

**TABLE III $R^2$ AND CORRELATION COEFFICIENT OF VOLUME CURVES**

| Target volume curve | $R^2$ | Correlation coefficient | RMSE |
|---|---|---|---|
| Combined T1T2 | 0.98 | 0.99 | 2.31 |
| Combined T2T3 | 0.96 | 0.98 | 5.50 |
| Combined T3T1 | 0.96 | 0.98 | 3.72 |
| Average | 0.97 | 0.98 | 3.84 |
| Standard deviation | 0.01 | 0.01 | 1.59 |

## 4. Discussion

The described process yielded multiple datasets, with which a first evaluation of our methodical approach was possible. To that end, the achieved results were viewed critically with regard to the initial hypothesis.

When placed correctly, the measurements correlated with B-mode imaging regarding a valuable position for a high likelihood of vessel coverage. Doppler measurements allowed for qualitative analysis of flow velocities and the resulting values correlated with the synchronically measured flow-values. This was verified visually as well as metrically via correlation calculation. However, the flexibility of the silicone tubes and the neck phantom tissue led to a different flow behavior when compared to the flow behavior inside the stiff flow-sensor. In all likelihood, this decreases the acceleration of velocity changes inside the tubes and the phantom, due to the pressure widening vessels which in turn releases pressure when flow velocity decreases. Effects of the inconsistent velocity profile along the vessel length could cause the curve differences shown in Fig. 4, Position 2. During moments of increasing flow volume, the volume flow increase is much steeper at the sensor than inside the phantom. Similarly, a decrease in flow and the corresponding pressure drop-off could be the cause of the longer-lasting flow inside the phantom, since the elastic components create an artificial pressurized tank (Windkessel) effect. While the minimal flow of both measurement approaches share temporal synchronicity, the flow inside the phantom never truly ceases, whereas flow-sensor values reach a minimal velocity of almost 0 m/s. Thus, the correlation calculations indicate a certain amount of similarity, however the average values are still below 0.9. The differences can originate in flow behavior between flow-sensor and phantom but also in the reflection combination due to the large SV, among others. The visual and metrical comparison allows for the conclusion that the proposed hardware can be used for the verification of measurement position goodness. For an exact flow calculation, the SV needs to be more precise. While positions 1 and 3 do not allow for any substantial vessel coverage via ultrasound, position 2 showed adequate results in B-Mode imaging and Doppler measurements which featured temporal correlation between flow curves.

However, the overall big SV of the initial measuring mode leads in all likelihood to an overlap of Doppler measurements along the vessel structure. A larger SV covers longer vessel distances with the ultrasound reaching different vessel segments at different times. This results in the summation of multiple frequency reflections in the final spectrogram, since the radial propagation of US waves does not result in a straight SV. Flow signals calculated from the envelope of the spectrograms consist therefore of reflections with multiple different Doppler angles, corrupting the results when substituting the Doppler angle with fixed transducer angles.

This has to be avoided, to allow for a correct volume calculation, which is why the initial mode was used to check for overall vessel coverage, but cannot be used for flow quantification. For a valid calculation, the SV has to be concentrated, preferably to a small volume inside the targeted vessel.

With the valid position for vessel coverage, the iterative acquisition of flow signals during periodical flow throughout the tissue depth was achieved with the step-wise increase of the SV depth. Results in Table II showed an increase in Doppler reflection power and the periodicity of the modified envelope curves when overlapping with SV ranges at the supposed vessel depth, measured via B-mode imaging. After subtraction of inner-probe distances between transducers and probe surface the distances of 15 mm to 17 mm, i.e. 12 mm and 14 mm after subtraction, matched the distance of vessel center and phantom surface. Vessel depth could therefore be approximated automatically under similar circumstances. Additionally, the switching of the transmitting transducer allowed for multiple measurement regimes that helped find the best ultrasonic emission angle. The combination of three different transducer angles or more in PW Doppler mode seems crucial in an emergency scenario. This way multiple measurement configurations allow for a more adaptive vessel localization and decrease susceptibility to non-ideal sensor placement, which can be observed in other scenarios [23].

The time for iterative measurements and transmitter switching is still a lot higher than the suggested five seconds [2], however the proposed system is in its infancy. The necessary time to find the vessel depth during resuscitation must be decreased drastically, which could be done by stripping the proposed method of the abstract calculation for spectrogram analysis and finding the best measurement parameters in the raw data. Achieving this would however require a viable database, which would have to be established with comparable measurements, using similar or the same sensor positions in relation to physiological landmarks and covering different patients' states multiple times. This would provide a basis for enhanced methods such as comparable analysis of the proposed calculations and raw signal behavior, e.g. with machine learning approaches.

The final volume calculation showed promising results when done with the small SV and the ultrasound configuration derived from the step-wise analysis. $R^2$, correlation coefficients, and RMSE show a strong correlation between them over the different flow states inside the laboratory setup. Inaccuracies might stem from the approximated flow volume calculation with certain simplification, e.g. the laminar flow inside the phantom or the area calculation based on a circle as crosscut. In general, the differences at each flow state are still considered significant, especially under what is considered close to ideal measurement conditions. However, the clinical acceptance of the overall system and specifically its maximal values for estimation errors need to be investigated further. Since there is no available dataset and no common consensus regarding the tolerance for flow estimation errors in non-invasive devices that target this specific physiological aspect, the evaluation of flow approximation tolerance is subject of studies in the future.

Overall, the methodical approach detailed in this publication could significantly contribute to the lack of information regarding carotid blood flow and possibly enable real-time resuscitation support. The chosen method distinguishes itself from alternatives by other groups, by enabling a more detailed automatic analysis regarding vessel location and direction. The specially designed sensor probe with fixed, non-symmetric ceramic angles provides a wider range of differently angled transmitter-receiver configuration. Incremental analysis of flow values inside small sample volumes can lead to the eventual identification of the best parametrization for a highly probable inner-vessel coverage. A smaller and mobile version would provide a measurement system that promotes stable, hands-free measurements, allows automatic parameterization and features a high grade of measurement repeatability, when the position of the sensor probe in relation to the neck of the volunteer and/or patient can be ensured. Future works could therefore focus on the implementation of a miniature device into a neck-stabilizer, similar to Laerdal Stifneck™ (Laerdal Medical GmbH, Stavanger, Norway). This would enable comparable measurements despite sensor detachment and reapplication and without the risk of skin trauma or irritation caused by adhesives. The method, the system and its integration with a Stiffneck are currently subject of a patent application (EP4487784A1) [24].

# 5. Conclusion

In conclusion, ultrasound hardware and analysis software were developed, and their capabilities tested in a laboratory environment. Tests aimed to evaluate whether the methods could be used to recognize flow in underlying tissue, to optimize ultrasound parameters and finally quantify measured flow inside a small SV. The results did not disprove the assumptions that the final demonstrator system can help in identifying valuable measurement positions for carotid PW Doppler and perform automatic parametrization during periodical flow activity. Further improvement of the hardware could be a miniature version with included analysis software that was described in this paper. The resulting system could be a considerable contribution to the understanding of carotid blood-flow during active CPR and provide users with more direct feedback on their resuscitation efforts. This would enable medical professionals and people without medical background to employ a new modality for the benefit of the patient, while adding to the necessary database for hemodynamic research. All in all, the results presented in this paper could enable further developments that could improve pre-clinical and clinical therapy based on guided resuscitation efforts.

**Acknowledgements**
The authors thank A. Eger for support on phantom casting processes, D. Gholami and T. Handel for support on hardware selection, and M. Nguyen Trong for software support.

**Author contributions**
Conceptualization, K.L., M.S., T.N., and R.F.; Formal analysis, M.S., J.S., G.S., M.U., and R.F.; Funding acquisition, K.L., G.S., T.N., and M.S.; Investigation, M.S., N.S. and R.F.; Methodology, M.U., J.S., R.F. and N.S.; Supervision, T.N., and K.L.; Writing – original draft, R.F., G.S., K.L., and M.S.; Writing – review & editing, all authors.

**Competing interests**
K.L., M.S., T.N., and R.F. are inventors on a patent application describing the methodical approach and sensor integration into a neck brace. M.S. has a financial interest in GAMPT mbH, a medical device company focused on the development of devices and materials for ultrasonic measurements. Research instrument support was provided by GAMPT mbH. The actuators and Doppler fluid used for the experiments in this publication were developed by GAMPT mbH. The authors declare that they have no other competing interests.

**Ethics declarations**
The research included neither human nor animal subjects.

**Funding**
This work was supported by the German Federal Ministry for Economic Affairs and Climate Action (Grant No. 16KN083328) and the German Federal Ministry of Education and Research (Grant No. 13GW0768B).

**Data and materials availability**
All data is available upon reasonable request to the corresponding author.

## Literature

1. Gräsner JT, Herlitz J, Tjelmeland IBM, Wnent J, Masterson S, Lilja G, et al. European Resuscitation Council Guidelines 2021: Epidemiology of cardiac arrest in Europe. Resuscitation. 2021 Apr;161:61–79.
2. Soar J, Bottiger BW, Carli P, et al. European Resuscitation Council Guidelines 2021: Adult advanced life support. Resuscitation 2021;161:115–51.
3. Kucewicz JC, Salcido DD, Adedipe AA, Truong K, Nichol G and Mourad PD. Towards a non-invasive cardiac arrest monitor: An in vivo pilot study. Resuscitation. 2019;134:76-80.
4. van Bel F, Roman C, Klautz RJ, Teitel DF and Rudolph AM. Relationship between brain blood-flow and carotid arterial flow in the sheep fetus. Pediatr Res. 1994;35:329-33.
5. Souchtchenko SS, Benner JP, Allen JL and Brady WJ. A review of chest compression interruptions during out-of-hospital cardiac arrest and strategies for the future. J Emerg Med. 2013;45:458-66.
6. Friess SH, Sutton RM, Bhalala U, Maltese MR, Naim MY, Bratinov G, Weiland TR, 3rd, Garuccio M, Nadkarni VM, Becker LB and Berg RA. Hemodynamic directed cardiopulmonary resuscitation improves short-term survival from ventricular fibrillation cardiac arrest. Crit Care Med. 2013;41:2698-704.
7. Naik R, Mandal I and Gorog DA. Scoring Systems to Predict Survival or Neurological Recovery after Out-of-hospital Cardiac Arrest. Eur Cardiol. 2022;17:e20.
8. Yilmaz G, Silcan M, Serin S, Caglar B, Erarslan O and Parlak I. A comparison of carotid doppler ultrasonography and capnography in evaluating the efficacy of CPR. Am J Emerg Med. 2018;36:1545-1549.
9. Adedipe AA, Fly DL, Schwitz SD, Jorgenson DB, Duric H, Sayre MR and Nichol G. Carotid Doppler blood-flow measurement during cardiopulmonary resuscitation is feasible: A first in man study. Resuscitation. 2015;96:121-5.
10. Zhao X, Wang S, Yuan W, Wu J and Li C. A new method to evaluate carotid blood-flow by continuous Doppler monitoring during cardiopulmonary resuscitation in a porcine model of cardiac arrest. Resuscitation. 2024;195:110092.
11. Lewis LM, Stothert JC, Jr., Gomez CR, Ruoff BE, Hall IS, Chandel B and Standeven J. A noninvasive method for monitoring cerebral perfusion during cardiopulmonary resuscitation. Journal of critical care. 1994;9:169-74.
12. Kenny JÉS, Munding CE, Eibl JK, Eibl AM, Long BF, Boyes A, u. a. A novel, hands-free ultrasound patch for continuous monitoring of quantitative Doppler in the carotid artery. Sci Rep. 8. April 2021;11(1):7780.
13. Faldaas BO, Nielsen EW, Storm BS, Lappegård KT, How OJ, Nilsen BA, u. a. Hands-free continuous carotid Doppler ultrasound for detection of the pulse during cardiac arrest in a porcine model. Resuscitation Plus. 20. Juni 2023;15:100412.
14. Song I, Yoon J, Kang J, Kim M, Jang WS, Shin NY, u. a. Design and Implementation of a New Wireless Carotid Neckband Doppler System with Wearable Ultrasound Sensors: Preliminary Results. Applied Sciences. Januar 2019;9(11):2202.
15. Huang H, Wu RS, Lin M, Xu S. Emerging Wearable Ultrasound Technology. IEEE Transactions on Ultrasonics, Ferroelectrics, and Frequency Control. Oktober 2023;71(7):713–29.
16. Wang F, Jin P, Feng Y, Fu J, Wang P, Liu X, u. a. Flexible Doppler ultrasound device for the monitoring of blood-flow velocity. Science Advances. 27. Oktober 2021;7(44):eabi9283.
17. Li L, Zhao L, Hassan R, Ren H. Review on Wearable System for Positioning Ultrasound Scanner. Machines. März 2023;11(3):325.

18. ACUSON Freestyle Wireless Ultrasound System [Internet]. [Last accessed 21. November 2024]. URL: https://www.siemens-healthineers.com/ultrasound/ultrasound-point-of-care/acuson-freestyle-ultrasound-machine
19. Clarius | Portable Pocket Handheld Ultrasound Scanners [Internet]. 2024 [Last accessed 21. November 2024]. URL: https://clarius.com/
20. Eibl J, Kenny JE, Magnin P, Eibl A. Systems and methods for automated fluid response measurement [Internet]. WO2017096487A1, 2017 [Last accessed 21. November 2024]. URL: https://patents.google.com/patent/WO2017096487A1/en?oq=US20170332995A1
21. Souza RM, Santos TQ, Oliveira DP, Souza RM, Alvarenga AV, Costa-Felix RPB. Standard operating procedure to prepare agar phantoms. J Phys: Conf Ser. Juli 2016;733:012044.
22. Li J, Zhang Y, Liu X, Liu P, Yin H, Liu DC. A Robust Automatic Ultrasound Spectral Envelope Estimation. Information. 5. Juni 2019;10(6):199.
23. McCann K, Holdgate A, Mahammad R, Waddington A. Accuracy of ECG electrode placement by emergency department clinicians. Emergency Medicine Australasia. Oktober 2007;19(5):442–8.
24. Fuchs R, Neumuth T, Eger A, Lenk K, Schultz M, Klaua R, et al. Ultrasonic Blood Flow Measurement Device for Blood Vessel [Internet]. EP4487784, 2025 [Last accessed 15. January 2025]. URL: https://www.freepatentsonline.com/EP4487784A1.html

## Supplementary

<u>Doppler Substitution</u>

Basis for substitution is the Doppler formula of the transducers $i$ and $j$, shown in equation (3) and equation (4) respectively.

$$\Delta f_i \;=\; f_0\,\frac{v}{c}\,(\,cos\,\alpha_{SE} \;+\; cos\,\alpha_{Ei}) \tag{3}$$

$$\Delta f_j \;=\; f_0\,\frac{v}{c}\,(\,cos\,\alpha_{SE} \;+\; cos\,\alpha_{Ej}) \tag{4}$$

$$\alpha_{Ei} \;=\; \alpha_{SE} + \beta_{i,SE} \tag{5}$$

$\Delta f_i$ and $\Delta f_j$ are the measured frequency differences, after ultrasound propagates through a matter with tissue propagation velocity $c$ once it reflected from an object that had the velocity $v$, with an initial frequency $f_0$. The Doppler angle is calculated by adding the angle between the transmitting transducer and flow velocity direction, $cos\,\alpha_{SE}$, and the angle at which the ultrasound reflects back to the receiver, $cos\,\alpha_{Ei}$ and $cos\,\alpha_{Ej}$. These angles can be broken down into the angle of the transmitter $\alpha_{SE}$ and the respective angle of incidence related to the normal of the probe surface, when ultrasound wave propagate from PEEK (c = 2570 m/s) to the agar mixture (c = 1540 m/s). Equation (5) shows this for transducer $i$

$$\frac{\Delta f_i}{\Delta f_j} = \frac{cos\,\alpha_{SE} + cos\,\left(\alpha_{SE} + \beta_{i,SE}\right)}{cos\,\alpha_{SE} + cos\,\left(\alpha_{SE} + \beta_{j,SE}\right)} \tag{6}$$

$$cos\,(\alpha_{SE} + \beta_{i,SE}) \;=\; cos\,(\alpha_{SE})\;\,cos\,(\beta_{i,SE}) \;-\; sin\,(\alpha_{SE})\,sin\,(\beta_{i,SE}) \tag{7}$$

When combining equation (3) and equation (4), equation (6) can be used to calculate the relation of frequency differences between the two transducers and ultimately the relation between the Doppler angles. The cosine calculation of $cos\,(\alpha_{SE} + \beta_{i,SE})$ can be split according to equation (7), to single out $\alpha_{SE}$ .

$$c_1 \;= \frac{\Delta f_i}{\Delta f_i}, c_2 = cos\,\left(\beta_{i,SE}\right), c_3 = sin\,\left(\beta_{i,SE}\right), c_4 = cos\,\left(\beta_{j,SE}\right), c_5 = sin\,(\beta_{j,SE}) \tag{8}$$

$$c_1[\,cos\,\alpha_{SE} \;+\; c_4 \times cos\,\alpha_{SE} \;-\; c_5 sin\,\alpha_{SE}\,] = \; cos\,\alpha_{SE} + \; c_2 cos\,\alpha_{SE} \;-\; c_3 sin\,\alpha_{SE} \tag{9}$$

$$(c_1 + c_1 \times \; c_4 - \; c_2 - 1)\; cos\,\alpha_{SE} \;=\; (c_1 \times \; c_5 - \; c_3)\; sin\,\alpha_{SE} \tag{10}$$

For readability the constants $c_1$ to $c_5$ are introduced in equation (8), which describe the known relation between the frequency measurements of both transducers and sine and cosine of the transducer incidence angles after refraction at the tissue border. Equation (7) can then be described as equation (9) and simplified to equation (10), where only known variables are used to describe the relation between $\cos\alpha_{SE}$ and $sin\,\alpha_{SE}$.

$$C = \frac{c_1 \times \, c_5 - c_3}{c_1 + c_1 \times \, c_4 - \, c_2 - 1} \;\text{ and } sin\,\alpha_{SE} \;=\; \sqrt{1 - \; cos^2 \alpha_{SE}} \tag{11}$$

$$cos\, \alpha_{SE} \;=\; C\sqrt{1 - cos^2\alpha_{SE}} \tag{12}$$

$$cos^2\, \alpha_{SE} \;=\; C^2(1 - cos^2\alpha_{SE}) \tag{13}$$

$$cos\, \alpha_{SE} \;=\; \sqrt{\frac{C^2}{1 + C^2}} \tag{14}$$

$$cos\left(\alpha_{SE} + \beta_{i,SE}\right) = cos(acos\,(cos\, \alpha_{SE}) + \beta_{i,SE}) \tag{15}$$

$$v \;= \frac{\Delta f_i}{f_0} \frac{c}{cos\, \alpha_{SE} + \; cos\, \alpha_{Ei}} \tag{16}$$

In equation (11) the relation between the sin and cosine of the transmitting transducers' angle $\alpha_{SE}$ is used as the constant $C$ while the sine is transferred to its' cosine equal. Through squaring of the equation (12), describing the transmitting transducers' angle relation in only cosine form, the equation (13) can be simplified to equation (14), which describes the sending transmitters' cosine angle form with only known constants from equation (8). The velocity of targets, i.e. erythrocytes in a blood stream or particles inside a Doppler fluid, can be calculated by reordering equation (3) to equation (16) and substituting $cos\, \alpha_{SE}$ according to equation (14) and $cos\, \alpha_{Ei}$ according to equation (15).